# Adaptive Control Plane Load Balancing in vSDN Enabled 5G Network

Deborsi Basu* *Graduate Student Member IEEE*, Uttam Ghosh† *Senior Member IEEE* and
Raja Datta‡ *Senior Member IEEE*
*G.S. Sanyal School of Telecommunication, Indian Institute of Technology, Kharagpur, India.
†Dept. of Electronics and Electrical Communication Engineering, Indian Institute of Technology, Kharagpur, India.
‡Dept. of Electrical Engineering and Computer Science, Vanderbilt University, Nashville, TN, USA.
deborsi.basu@iitkgp.ac.in*, ghosh.uttam@ieee.org†, rajadatta@ece.iitkgp.ac.in‡

## I. INTRODUCTION

Standardization of the future 5G networks are still under construction and researchers are intensively working into this field [1], [2]. However, the evolution of the cellular technology is ever expanding and will flow beyond 5G as well, soon after the standardization has been made successfully. The speculations based on 6G networks are already in the market and people have already started thinking beyond 5G network architectures and technological developments. The fundamental aim of all such developments always remains the same, which is to satisfy the end user demands in all possible way. Everyday network servers are being loaded with hundreds of new applications and services & maximum of them are redundant in nature, creating unnecessary load overhead on the network. Keeping demand threshold restriction in mind the TSPs (Telecomm. Service Providers) are giving intensive care to server load distribution for handling heavy network functions without any service disruption. The integration of SDN & NFV give the tenants an upgraded open platform to manage these complex issues by providing a virtualized network service plane with dynamic and programmable network slicing architectures. The openness of networks brings major network scalability problems like CPP & HPP (Controller Placement Problem & Hypervisor Placement Problem). Network Hypervisors are virtual entities to be placed in between Control Plane and Data Plane to allow multiple tenants to use the network resources and update systems simultaneously [3–5]. Due to centralized architecture SDN faces huge traffic load at the control plane which degrades network performances [6], [7]. Through our work we have proposed an intelligent solution of Control Plane load balancing of Joint CPP-HPP using vSDN concept, that transforms non-virtualized networks to virtualized networks. We have used real network topology to validate our approach by defining four latency objectives and finding out the potential Controller-Hypervisor (C-H) positions that generates minimal load on C-H planes. This approach can also be applied in solving similar other critical localization problems like service chain mapping in 5G-NR, baseband unit deployment in 5G C-RAN etc.

## II. RELATED WORK & MOTIVATION

The CPP was first framed by Heller *et al.* in [8] where they have tried to find out the number and potential positions of the controller inside a given 100 possible WAN (Wide Area Network) topology. The introduction of Hypervisor plane or H-plane has been done inside the CPP by Blenk *et al.* in [9]. They formulated a Hypervisor Placement Problem (HPP) in a vSDN network while fixing the controller positions at each vSDN. Furthermore, Killi *et al.* in [10] have shown that reduction of worst-case latency can be done by fixing the hypervisor location at H-plane and optimizing the controller location at C-plane. Furthermore, it is shown that the Joint HPP-CPP models at physical networks and virtual networks respectively are even more efficient in latency reduction. The flow-paths which are terminated at H-plane via C-plane makes unnecessary load at H-plane. They are responsible for higher network processing delay. If the path-flow can be reverted back to C-plane only then the load at H-plane can be reduced significantly. The flow signals which reach the C-plane first through the shortest paths are being processed at C-plane itself. This conceptualization motivates us to contribute further in this Joint HPP-CPP load balancing model & our major contributions have been explained next.

**Contributions:** In this work we have initiated the process of adaptive load balancing by solving Joint HPP-CPP in the context of future 5G networks. More precisely the contributions have been explained below:

- We have started the Joint C-plane and H-place controller and hypervisor placement using a real network topology & the flexibility to select both their positioning in the planes. We have defined an MILP based analytical modeling on graphical mapping to solve the problems based on four latency objectives.
- After getting the proper positions of the physical entities at both the C and H planes based on the latency objectives, we have found out the optimum propagation paths from the user end to the controller using a bottom up approach. Our proposed RPF (Reverse Path-Flow) algorithm suitably found out the optimal paths from the Data Plane physical nodes to the controller through which the PACKET-IN messages flow. It can be seen that all the paths reaching to the C-plane are not necessarily routed through the H-plane. We restrict the unnecessary in flow of the messages at the H-plane.
- Finally, we discarded the extra paths to reduce the load at the H-plane and C-plane overheads as well. As H-plane has been used by multiple tenants to bring their network services, it is extremely necessary to keep the load at the H-plane as low as possible to give high processing efficiency to the service messages. The C-H plane load & propagation delay reduction of Service Requests are the key enablers to experience ULL network services to which the future 5G & 6G networks are aiming [11].

## III. SYSTEM MODEL & PROBLEM FORMULATION

### A. System Model

We have considered a real network topology instance of AT&T North America and apply an analytical graphical modeling to make the Load and Latency aware Joint HPP-CPP formulation. We have compared our simulation results with the existing CPP-HPP models [9], [10] to show the efficiency of our proposed approach. For simplicity we have considered two terminologies where Joint CPP-HPP is represented by **JHCPP** and Load and Latency aware model has been represented by **opJHCPP** (Open JHCPP- Open Joint Hypervisor-Controller Placement Problem) and the same terms have been maintained throughout the paper.

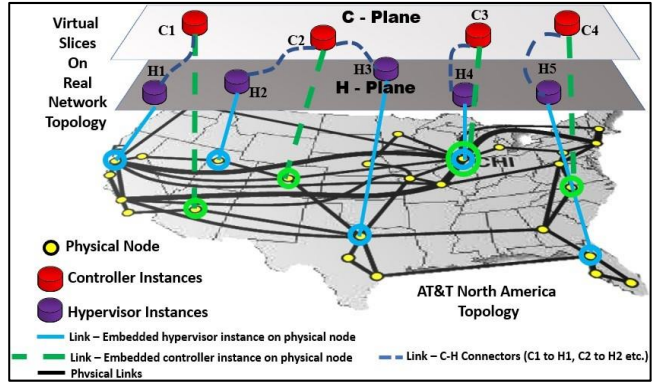

Fig. 1: The controller and hypervisor positions based on latency matrix optimization - A Generalized Case [13]

The network has been considered as a graph denoted by $\mathcal{G} = V_p \rightarrow physical\ node \cup E_{link} \rightarrow bi-directional\ links$. Physical Nodes $p_{node} \in V$ can consist zero, one or multiple virtual hypervisor or controller instances ($h_{node} \in H_{node}$ & $c_{node} \in C_{node} \subseteq V_p$). $H_{node}$ & $C_{node}$ are the potential positions of the hypervisors and controller nodes respectively. The maximum number of hypervisor and controller instances for a single virtual network is $H_{inst}$ and $C_{inst}$ respectively. The cost of the min. distance path from a physical vSDN node to the controller through hypervisor for a particular service demand towards the controller is given by $\psi_{v_n,\ d,\ h_{node},\ c_{node}}$. The path with source, intermediate and destination at same node costs zero.

### B. Problem Formulation

Now the following **opJHCPP** Latency matrices are explained below which are evaluated and compared with other existing cases mentioned in [9], [10] to show the effectiveness of our proposed approach.

*1) Worst Case Latency:* This is used to minimize the longest path delay among all vSDNs.

$$min\ (\mathbb{W}) \qquad (1)$$

Given the following: The worst-case latency or Maximum latency case can be incorporated in our model by adding a continuous variable constraint in the equation (2) given by $\mathbb{W}$.

$$L_{worst} = \left(\max_{v_n \in V_{net},\ d\ \in\ D_{v_n}}\right)\left(\sum_{h_{node} \in H_{node},\ c_{node} \in C_{node}}(Y_{v_n,\ d,\ h_{node},\ c_{node}} \times \psi_{v_n,\ d,\ h_{node},\ c_{node}})\right) \qquad (2)$$

Where,

$$\sum_{h_{node} \in H_{node},\ c_{node} \in C_{node}}(Y_{v_n,\ d,\ h_{node},\ c_{node}} \times \psi_{v_n,\ d,\ h_{node},\ c_{node}}) \leq \mathbb{W}\ \forall\ v_n \in V_{net},\forall\ d\ \in\ D_{v_n} \qquad (3)$$

*2) Minimum of Average Latency:* The minimum overall average latency demands among all vSDNs.

$$min\ (L_{avg}) \qquad (4)$$

Given the following:

$$\frac{1}{\sum_{v_n \in V_{net}}|D_{v_n}|} \sum_{v_n \in V_{net}} \sum_{d\ \in\ D_{v_n}} \sum_{h_{node} \in H_{node},\ c_{node} \in C_{node}}(Y_{v_n,\ d,\ h_{node},\ c_{node}} \times \psi_{v_n,\ d,\ h_{node},\ c_{node}}) \qquad (5)$$

*3) Avg-Maximum Latency:* It selects all the worst-case latencies from each vSDN and then try to minimize the average value of the respective latencies.

$$min\ \left(\frac{1}{|V_{net}|} \sum_{v_n \in V_{net}} \mathbb{W}_{v_n}\right) \qquad (6)$$

Given the following: Here the continuous variable $\mathbb{W}_{v_n}$ represents the worst-case latencies of individual virtual networks $v_n$ given by equation (2).

$$L_{avg-max} = \frac{1}{|V_{net}|}\left(\max_{d\ \in\ D_{v_n}}\right) \sum_{h_{node} \in H_{node},\ c_{node} \in C_{node}}(Y_{v_n,\ d,\ h_{node},\ c_{node}} \times \psi_{v_n,\ d,\ h_{node},\ c_{node}}) \qquad (7)$$

Where,

$$\sum_{h_{node} \in H_{node}, \ c_{node} \in C_{node}} (Y_{v_n, \ d, \ h_{node}, \ c_{node}} \times \psi_{v_n, \ d, \ h_{node}, \ c_{node}}) \leq \mathbb{W} \ \forall \ v_n \in V_{net}, \forall \ d \in D_{v_n} \quad (8)$$

*4) Max-Average Latency:* It takes all the average values from each vSDN and then aim to find the minimum among all maximums.

$$\min(\mathbb{W}) \quad (9)$$

Given the following: Similarly, here also $\mathbb{W}$ is a continuous variable for max-avg latency of individual vSDNs.

$$L_{max-avg} = \left(\max_{v_n \in V_{net}}\right) \frac{1}{|D_{v_n}|} \sum_{d \in D_{v_n}} \sum_{h_{node} \in H_{node}, \ c_{node} \in C_{node}} (Y_{v_n, \ d, \ h_{node}, \ c_{node}} \times \psi_{v_n, \ d, \ h_{node}, \ c_{node}}) \quad (10)$$

Where,

$$\frac{1}{|D_{v_n}|} \sum_{d \in D_{v_n}} \sum_{h_{node} \in H_{node}, \ c_{node} \in C_{node}} (Y_{v_n, \ d, \ h_{node}, \ c_{node}} \times \psi_{v_n, \ d, \ h_{node}, \ c_{node}}) \leq \mathbb{W} \ \forall \ v_n \in V_{net} \quad (11)$$

## IV. SOLUTION & RESULT ANALYSIS

### A. Algorithmic Solution

We have used a Greedy Algorithmic approach to solve the above opJHCPP problem using the Reverse path-flow concept.

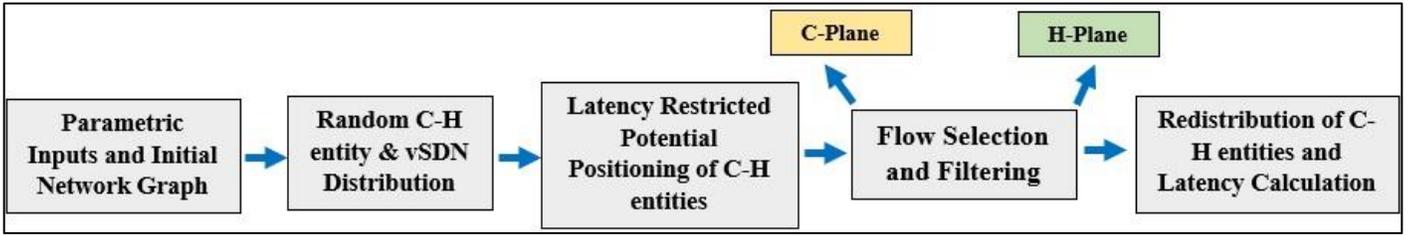

Fig. 2: The Sequence Flow Diagram of Algorithmic Solution

### B. Result Evaluation

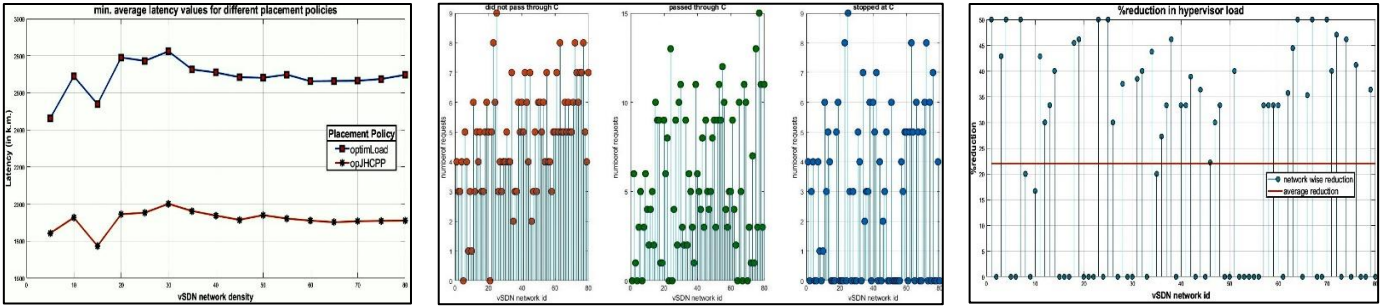

(a) min. avg. latency comparison    (b) test case signals-block and pass    (c) test case H-plane load reduction

Fig. 3: Comparative load reduction using opJHCPP with latency trade-off

For calculating the percentage reduction in hypervisor load it is necessary to compute the distribution pattern of requests based on whether they were blocked by the controller, passed through the controller or did not pass through the controller in an attempt to reach the network hypervisor. For a particular $c_{node}$-$h_{node}$ pair let's denote the number of requests which did not pass through the controller in order to reach the hypervisor by *dptc*, number of requests which were forwarded from the controller to the hypervisor by *cp* and the number of requests which were blocked at the controller by *cs*. Then the reduction $r_{c_{node}, \ h_{node}}$ is given by:

$$r_{c_{node}, h_{node}} = \frac{cs}{cs + cp + dptc} \quad (12)$$

The distributions of requests for different controller and hypervisor combinations for $|M| = 80$ (no. of vSDN networks) with respect to the vSDN network ids are obtained. (i,j) refers to controller position C(i) and hypervisor position H(j). Here C = 3,7,10,23 and H = 2,5,15,19 are converging potential positions we get after 100 iterations. It can be seen clearly that the controller and hypervisor positions best suited for maximum hypervisor load reduction are not favorable for achieving optimum latency benefits for the latency objectives. Hence, there is a trade-off between minimizing hypervisor load and minimizing network latency. It can be explained from the fact that minimum hypervisor load in case of a single controller and single hypervisor system (as considered in this case) means that most of the requests are blocked by the controller and do not reach the hypervisor which itself means that such controller and hypervisor positions are far apart, so for those requests which do reach the hypervisor, the latencies of propagation are maximum. One important fact to be noted is that, the heavy messages are processed at the Controller itself and restricted for further transmission to H-plane only if their *C-plane processing time ≤ (2 \* (C plane to H plane propagation delay) + H-plane processing time)*. The trade-off in latencies can be minimized by placing multiple hypervisors and controllers in the networks which we may consider in our future works.

## V. Conclusion & Future Works

Our work provides a comprehensive adaptive load balancing approach by properly positioning the controller and hypervisor entities at H-C planes according to their potential locations. Here, we have kept the network service latency under suitable tolerance limit. Apart from our topology (AT&T North America), the proposed approach can be applied to other network topologies as well by changing the parametric values according to the system requirements. We have illustrated the effect of different evaluation setups on the outcome of all four latency metrics. A C++ algorithm-based framework has been implemented using MATLAB R2020a to simulate, & solve the problem, and also analyze the results. In future, we will be targeting some interesting facts like AI (Artificial Intelligence) driven task offloading in between Hypervisor plane and Control plane for optimal resource utilization. We will be focusing towards network slicing based edge computing approach to allocate the additional network functions for optimum QoE for end-users.

## Authors' Bio

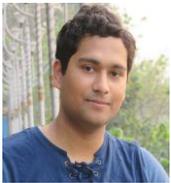

**Deborsi Basu** is pursuing his Ph.D from G.S Sanyal School of Telecommunication, Indian Institute of Technology, Kharagpur, India with the joint collaboration of Dept. of Electrical Engineering and Computer Science, University of Vanderbilt, Nashville, Tennessee, USA. He has completed his M.Tech from Kalyani Government Engineering College, Kalyani, West Bengal, India in Dept. of ECE in 2018 and B.Tech from Heritage Institute of Technology, Kolkata, West Bengal, India in Dept. of ECE in 2016. He is a Graduate Student Member of IEEE. His current research areas are Software Defined Networking, Network Function Virtualization, Network Slicing in 5G & NextGen Wireless Communication Networks, OpenFlow etc.

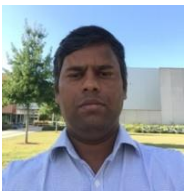

**Uttam Ghosh** joined as an Assistant Professor of the Practice in the Dept. of Electrical Engineering and Computer Science in January 2018 at Vanderbilt University. Dr. Ghosh obtained his PhD in Electronics and Electrical Engineering from Indian Institute of Technology Kharagpur, India in 2013, and has Post-doctoral experience at the University of Illinois in Urbana-Champaign, Fordham University, and Tennessee State University. His main research interests include Cybersecurity, Computer Networks, Wireless Networks, Information Centric Networking and Software-Defined Networking. He is actively working with VECTOR Vanderbilt University. Designer of TennSMART consortium to accelerate "Intelligent Mobility" in Tennessee State University.

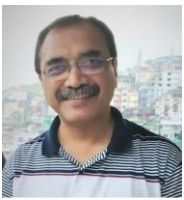

**Raja Datta** is the Head of the department at G.S Sanyal School of Telecommunications at Indian Institute of Technology (IIT) Kharagpur. He also is the Professor In-charge of Technology Telecom Center. Prof. Datta is a senior member of IEEE. He has produced a number of PhD and MS students in area of Communication Networks. His research areas include computer communication networks, mobile ad-hoc and sensor networks, optical WDM networks, inter planetary networks, computer architecture, distributed systems etc.